# Dispersion of the superconducting spin resonance in underdoped and antiferromagnetic Ba(Fe$_{1-x}$Co$_x$)$_2$As$_2$

D. K. Pratt[1], A. Kreyssig[1], S. Nandi[1], N. Ni[1], A. Thaler[1], M. D. Lumsden[2], W. Tian[1], J. L. Zarestky[1], S. L. Bud'ko[1], P. C. Canfield[1], A. I. Goldman[1], and R. J. McQueeney[1]

[1]*Department of Physics and Astronomy and Ames Laboratory, Iowa State University, Ames, IA 50011 USA*
[2]*Oak Ridge National Laboratory, Oak Ridge, TN 37831 USA*

*Abstract:* Inelastic neutron scattering measurements have been performed on underdoped Ba(Fe$_{1-x}$Co$_x$)$_2$As$_2$ ($x = 4.7\%$) where superconductivity and long-range antiferromagnetic (AFM) order coexist. The broad magnetic spectrum found in the normal state develops into a magnetic resonance feature below $T_C$ that has appreciable dispersion along *c*-axis with a bandwidth of 3-4 meV. This is in contrast to the optimally doped $x = 8.0\%$ composition, with no long-range AFM order, where the resonance exhibits a much weaker dispersion [see Lumsden et al. Phys. Rev. Lett. **102**, 107005 (2009)]. The results suggest that the resonance dispersion arises from interlayer spin correlations present in the AFM ordered state.

Similar to many other unconventional superconductors (SC), magnetism and SC are intimately linked in the newly discovered iron pnictide compounds. In both electron doped BaFe$_2$As$_2$ materials, SC appears only after the antiferromagnetic (AFM) ordering observed in the parent compound is suppressed.[1-4] However, the suppression of AFM order need not be complete, and both SC and long-range AFM order can coexist in so-called underdoped (UD) regions of the phase diagram.[1,4] For these UD compositions, it has been shown that SC and static AFM order are in competition, being characterized by a substantial reduction of AFM order parameter below $T_C$.[5,6] Another indication of the relationship between magnetism and SC comes from the observation of a magnetic resonance mode appearing below $T_C$ by inelastic neutron scattering. The resonance has been observed in optimally doped (OD) compositions (defined as having near maximum $T_C$ with no long-range AFM order) [7-10] as well as UD compounds.[5,6] In both doping limits, the resonance appears near $\mathbf{Q}_{AFM}$, the wavevector of the ordered AFM structure, and is sharply peaked at $\mathbf{Q}_{AFM}$ for momenta in the Fe layers. On the other hand, the resonance varies weakly along the $c$-axis (perpendicular to the Fe layers) for optimal Co-doping [7] and Ni-doping,[8] suggesting nearly two-dimensional (2D) behavior. Similar to the cuprates,[11] the energy of the resonance mode in OD iron pnictide compounds is in the range of 4-5$k_BT_C$ and can therefore be associated with the SC gap energy.[12]

In the UD compositions, where AFM order persists in the SC state, the effect of the AFM order on the resonance dispersion and relationship between the resonance and spin wave excitations must also be considered. Here we show that the resonance in UD Ba(Fe$_{1-x}$Co$_x$)$_2$As$_2$ with $x = 4.7\%$ ($T_N = 47$ K, $T_C = 17$ K) disperses quite strongly along the $c$-axis (within an energy window of 4-8 meV). This is compared to the nearly dispersionless resonance found at ~9 meV in OD composition ($x = 8.0\%$, $T_C = 22$ K), indicating that both the energy and bandwidth of the resonance are composition dependent.[7] The results suggest that AFM order leads to resonance dispersion and we show that the resonance bears some similarity to the AFM spin waves themselves.

Inelastic neutron scattering measurements were performed on the HB3 spectrometer at the High Flux Isotope Reactor at Oak Ridge National Laboratory on single-crystals of Ba(Fe$_{1-x}$Co$_x$)$_2$As$_2$ with $x = 4.7\%$. The sample consists of 9 co-aligned crystals with a total mass of 1.88 grams and a total mosaic width of 1.5 degrees. All samples were grown under identical conditions with tetragonal-orthorhombic transition ($T_S = 60$ K), Neel transition ($T_N = 47$ K) and SC transition temperatures ($T_C = 17$ K), consistent with crystals used in previous studies.[2,10] Although the measurements were made below $T_S$ and the sample is orthorhombic, in what follows we describe the scattering vector relative to the high temperature tetragonal (*I4/mmm*) cell. The sample was aligned in the [*H H L*] plane and mounted in a closed-cycle refrigerator for low temperature studies. Measurements were performed on HB3 spectrometer with 48'-60'-80'-120' collimation and a fixed final energy of $E_f = 14.7$ meV. A pyrolytic graphite (PG) monochromator and analyzer were employed. One PG filter was used after the sample for inelastic measurements, while two filters were used for elastic measurements to reduce the signal from higher order harmonics.

Figure 1 summarizes several features of the neutron intensity [$I(\mathbf{Q},\omega)$] above and below $T_C$ in the ordered AFM state. In Fig. 1(a), the energy dependence of the scattering is shown above $T_C$ ($T = 25$ K) and below $T_C$ ($T = 5$ K) at $\mathbf{Q}_{\text{AFM}} = (½ ½ 1)$. As reported previously,[5] the data shows a resonance feature at $\mathbf{Q}_{\text{AFM}}$ that arises from the redistribution of magnetic intensity from low energies to high energies below $T_C$. Fig. 1(a) also shows estimates of the non-magnetic background [$C(\omega)$] at both temperatures as obtained from scans at $\mathbf{Q} = (0.35\ 0.35\ 1)$ and $(0.65$ 0.65 1) which are far from magnetic intensity centered $\mathbf{Q}_{AFM}$ and display featureless energy response.

These data can be used to estimate the imaginary part of the magnetic susceptibility at $\mathbf{Q}_{AFM}$ using the equation below,

$$\chi''(\mathbf{Q}_{AFM},\omega) = [I(\mathbf{Q}_{AFM},\omega) - C(\omega)](1 - e^{-\hbar\omega/kT}) \quad (1)$$

as shown in Fig. 1(b). The linear energy dependence of the normal state susceptibility for $\hbar\omega < 6$ meV suggests gapless excitations although we cannot ascertain whether a small gap exists below 2 meV due to finite instrumental resolution. A comparison of $\chi''$ at 25 K and 5 K shows that the resonance exhibits an onset at 4 meV, a peak near 5 meV, and a long tail extending up to 10 meV.

Fig 1(c) and 1(d) explore the $\mathbf{Q}$-dependence of the magnetic scattering, showing constant energy scans at 5, 7, and 10 meV along the [$H\ H\ 0$] and [$0\ 0\ L$] directions through $\mathbf{Q}_{AFM}$. As expected for the ordered AFM state, the normal state excitations along [$H\ H\ 0$] are sharply peaked at (½ ½ 1) and appear to be consistent with the steep spin wave dispersion observed in parent compounds. The normal state lineshapes are much broader along the $L$-direction than the corresponding [$H\ H\ 0$] scans due to the relative weakness of the interlayer exchange. The normal state spin excitations above $T_C$ were fit using a damped spin wave model convoluted with the instrumental resolution function [lines in Figs. 1(c) and 1(d)]. The damped spin wave response function used for analysis is

$$\chi''(\mathbf{q},\omega) = \frac{\Gamma\omega}{(\omega^2 - \omega_\mathbf{q}^2)^2 + \Gamma^2\omega^2} \quad (2)$$

$$\omega_\mathbf{q} = \sqrt{v_{ab}^2(q_x^2 + q_y^2) + v_c^2 q_z^2 + E_g^2)} \quad (3)$$

where $v_{ab}$ and $v_c$ are the in-plane and interplane spin wave velocities, $E_g$ is an anisotropy gap, and $\Gamma$ is a damping parameter and the wavevector **q** is defined relative to the magnetic Brillouin zone center at $\mathbf{Q}_{AFM} = (½ ½ 1)$. Fits to normal state spin waves in the [H H 0] and [0 0 L] directions through $\mathbf{Q}_{AFM}$, shown in Fig. 1, yielded spin wave velocities of $v_{ab} \geq 123$ meV·Å and $v_c = 43 \pm 9$ meV·Å. The damping and the anisotropy gap parameters were obtained by fits to both the linear dispersion in Eq. (3) as well as a more general Heisenberg model (not shown) and were found to be in the range of $\Gamma = 8$ -12 meV and $E_g = 7$- 9 meV. Thus, despite an apparent finite value of the anisotropy gap, spectral weight persists to the lowest measurable energies due to significant damping [Fig. 1(b)]. However, we note that the damping and anisotropy gap parameters depend sensitively on estimates of the non-magnetic background. The fits to a Heisenberg model were generally consistent with the published results for $x = 4.0\%$.[6]

Below $T_C$, Figs. 1(c) and 1(d) also show the **Q**-depenence of the resonance which appears as a change in the intensity of the magnetic scattering below $T_C$. Similar to other iron pnictides, the SC resonance is sharply defined for wavevectors in the Fe-layer near $\mathbf{Q}_{AFM}$. The effect of SC on the spin excitations propagating along $L$ is much more interesting. Measurements of the $L$-dependence at the resonance peak energy at 5 meV show that it is narrowly defined at $L = 1$, similar to reports for $x = 4.0\%$.[6] At a slightly higher energy of 7 meV, the intensity of the resonance appears to broaden or shift away from $L = 1$. At 10 meV, the resonance intensity has weakened considerably and can be found only near the magnetic Brillouin zone boundary at $L = 0$ or 2. This is very different from the $L$-dependence observed in the OD compound with $x = 8.0\%$, as determined by Lumsden *et al*. in Ref . [7], where the intensity of the resonance peak (at 10 meV) is broadly distributed along $L$ (Fig. 1, inset).

These results suggest that the resonance observed in UD compositions has dispersion along the $L$-direction. This dispersion is seen more clearly in energy scans measured at $\mathbf{Q} = (½\ ½\ L)$ for several values of $L$ and temperatures above and below $T_C$, as shown in Fig. 2(a). As $L$ is increased away from $\mathbf{Q}_{AFM}$, the resonance intensity shifts to higher energies with $L$ and weakens, being nearly absent at $L = 2$. Fig. 2(b) compares the difference of intensities measured at 25 K and 5 K and $L = 0, 1/2$, and 1 as compared to the OD from Lumsden *et al.* (Ref. [7]), illustrating the distinctive behavior of compositions with and without AFM order. Fig. 2(c) compares the $L$-dispersion of the resonance peak for UD and OD samples. The UD resonance peak disperses from 5 meV at $\mathbf{Q}_{AFM} = (½\ ½\ 1)$ to 8 meV at the zone boundary whereas the resonance of the OD sample remains nearly dispersionless in the range of 8-9 meV.

The resonance dispersion can be fit to an empirical function $\Omega(L) = \Omega_0 + W|\cos(\pi L/2)|$ where the $\Omega_0$ is the energy of the resonance at $\mathbf{Q}_{AFM}$ and $W$ is the bandwidth. For $x = 4.7\%$, $W = 3$ meV and $W/\Omega_0 = 0.6$, whereas $W/\Omega_0 < 0.1$ for $x = 8.0\%$, as shown in Fig. 2(c). This change in the relative bandwidth with doping suggests that the magnetic resonance is a three-dimensional (3D) feature when AFM and SC coexist and evolves to a 2D feature upon the loss of magnetic order.

We now discuss the possible origin of the resonance dispersion in the AFM ordered state. The magnetic resonance in iron pnictide SC has been interpreted in the context of a spin exciton model [13,14] where the normal state spin fluctuations arising from quasiparticle excitations become gapped below $T_C$. Within the random-phase approximation (RPA), the dynamical magnetic susceptibility in the SC state is given by

$$\chi(\mathbf{q},\omega) = \frac{\chi_0(\mathbf{q},\omega)}{1 - J(\mathbf{q})\chi_0(\mathbf{q},\omega)} \tag{4}$$

where $\chi_0$ is the non-interacting dynamical susceptibility in the SC state and $J$ is an interaction parameter. Due to the coherence factors in the SC state, a magnetic resonance appears in $\chi''$ at the energy where $\chi_0 = 1/J$ with strong enhancement when the SC order parameter has sign-reversing symmetry $\Delta_{\mathbf{k}} = -\Delta_{\mathbf{k}+\mathbf{Q}_{AFM}}$.[13,14] For an s-wave gap with this property ($s_{+-}$), the RPA theory predicts that $\Omega_0 \approx 1.4\Delta$ for 2D spin fluctuations at $\mathbf{Q}_{AFM}$. This relationship holds for $x = 8\%$,[7] consistent with both the spin exciton picture and the $s_{+-}$ symmetry of SC order parameter.

In the presence of AFM order, the interlayer exchange interaction will give rise to dispersion of the normal state (spin wave) excitations. The dispersion is given by $\hbar\omega(q_z) \sim J(0) - J(q_z)$ where $J(q_z)$ decreases with increasing $q_z$, as shown in Figs. 3(c) and 3(d). Within RPA, the pole giving rise to the resonance at $\chi_0 = 1/J$ will then disperse upwards in energy upon moving away from $\mathbf{Q}_{AFM}$ in the $L$-direction with the maximum energy bounded by the SC gap, $\Omega(L) \leq 2\Delta$. Even without AFM long-range order, the presence of pronounced short-ranged spin correlations along $L$ has been used to explain the weaker resonance dispersion ($W/\Omega_0 = 0.26$) observed in Ba(Fe$_{1-x}$Ni$_x$)$_2$As$_2$.[8]

Another hallmark of unconventional SC in quasi-2D antiferromagnets is that the ratio $\Omega_0/k_B T_C$ is usually found in the range from 4-5.[15] For $x = 4.7\%$, this ratio is only 3.5, which is somewhat smaller than expected. Even more surprising is that resonance observed in another UD composition with $x = 4.0\%$ ($\Omega_0 \approx 4.5$ meV) [6] has a similar energy as $x = 4.7\%$. This is true despite the fact that $T_C = 11$ K for $x = 4.0\%$, whereas $T_C = 17$ K for $x = 4.7\%$ indicating that there is no scaling between $\Omega_0$ and $T_C$. It is interesting to note that resonance dispersion is observed in UPd$_2$Al$_3$ ($\Omega_0/kT_C = 2.8$) where SC also appears within an AFM ordered state ($T_N = 17$ K, $T_C = 2$ K).[16,17] CeCoIn$_5$ is not magnetically ordered, however, strong interlayer spin correlations exist resulting in an $L$-dependent resonance where $\Omega_0/kT_C = 3$.[18]

The near constant value of $\Omega_0$ in UD 4.0% and 4.7% hints that $\Omega_0$ in AFM ordered systems is influenced by another energy scale, such as the normal state spin wave dispersion, anisotropy gap, and/or Landau damping. Figure 3(a) shows a contour plots of the normal state susceptibility as a function of $L$ and $\hbar\omega$ with the fitted spin wave dispersion superposed. The zone boundary spin excitation is estimated to be 20 meV with substantial damping $\Gamma = 10$ meV [the calculated damped spin wave susceptibility is shown in Fig. 3(b)]. Fig 3(c) shows the measured susceptibility below $T_C$ with both the resonance and spin wave dispersions superposed on the image. Fig. 3(d) shows the resonance dispersion as the difference of the normal and SC state susceptibilties. The fairly low energy spin wave dispersion along $L$, combined with large damping, is suggestive of the magnon scenario for the resonance [19] where the SC gap acts to reduce the Landau damping and the subsequent sharpening of spin wave modes near or below $2\Delta$ yields a resonance-like feature. The magnon scenario has been used to describe the resonance in the electron-doped cuprates,[20], UPd$_2$Al$_3$,[21] and CeCoIn$_5$.[22]

In summary, the weakly dispersive magnetic resonance observed in the superconducting state of OD, paramagnetic, Ba(Fe$_{1-x}$Co$_x$)$_2$As$_2$ is quasi-2D, but develops significant dispersion along $L$ for UD compositions where AFM and SC coexist (i.e. becomes more 3D). The resonance energy at $\mathbf{Q}_{AFM}$ does not appear to scale with $T_C$ suggesting that the resonance may not have a simple or universal relationship to the SC gap when AFM order exists. The AFM order can be considered as a spin-density wave (SDW) that is stabilized by the gapping of the Fermi surface at $\mathbf{Q}_{AFM}$. In the scenario that the SDW and SC phases compete as the two gaps vie for the same electrons on the Fermi surface,[23] the interplay of SC and spin excitations, and consequently the resonance, is demonstrably more complex in the presence of AFM order.

*Acknowledgments*. The authors would like to thank J. Schmalian, V. P. Antropov, M. R. Norman, R. M. Fernandes, and I. Eremin for valuable comments. Work at the Ames Laboratory is supported by the U.S. Department of Energy Office of Science under Contract No. DE-AC02 07CH11358. Work at Oak Ridge National Laboratory is supported by the Scientific User Facilities Division, Office of Basic Energy Sciences.

FIGURE CAPTIONS

FIG. 1. (a) The raw neutron intensity $I(\mathbf{Q},\omega)$ for $x = 4.7\%$ including background measurement for energy scans at $\mathbf{Q}_{AFM} = (½ ½ 1)$ both above (25 K, open circles) and below $T_C$ (5 K, filled circles). The arrow shows the location of the resonance. (b) Energy dependence of $\chi''(\mathbf{Q}_{AFM},\omega)$ at 5 and 25K. The solid line is a fit to spin waves described in the text. (c) $(H\,H\,0)$ scans and (d) $(0\,0\,L)$ scans through $(½ ½ 1)$ at 5, 7,and 10 meV. In (c), and (d), solid lines represent fits to the spin wave model in the normal state. The inset to shows $(0\,0\,L)$-scans for $x = 8.0\%$ both above (30 K, open triangles) and below $T_C$ (10 K, filled triangles) at an energy transfer of 9.5 meV (close to the resonance peak) taken from Lumsden *et al.* Ref. [7].

FIG. 2. (a) The dispersion of the resonance is shown by constant-$\mathbf{Q}$ energy scans at $(½ ½ L)$ for $x = 4.7\%$ for several values of $L$ at temperatures above and below $T_C$. The dark grey shading highlights regions of increased intensity related to the resonance while the light grey shading highlights the loss of intensity. (b) Difference of scattering intensity between temperatures above and below $T_C$ for both $x = 4.7\%$ (open circles) and 8.0% (filled triangles) at $L = 0, 1/2$, and 1. (c) A comparison of the dispersion of the resonance peak energies along the $L$-direction for $x = 4.7\%$ and 8.0%. The $x = 8.0\%$ data are taken from Ref. [7].

FIG. 3. Contour plots of the magnetic susceptibility as a function of $\hbar\omega$ and $L = 1 - 2$. The contour plots have been reflected through $L = 1$. (a) Measured data at 25 K. The line shows the fitted normal state spin wave dispersion. (b) Normal state damped spin wave fitting results. (c) Measured data below $T_C$ at 5 K. The solid line is the normal state spin wave dispersion from panel (a) and the combined line and square symbols are the fitted resonance dispersion. (d) The measured resonance susceptibility obtained from the difference of the data at 25 K [panel (a)] and 5 K [panel (c)].

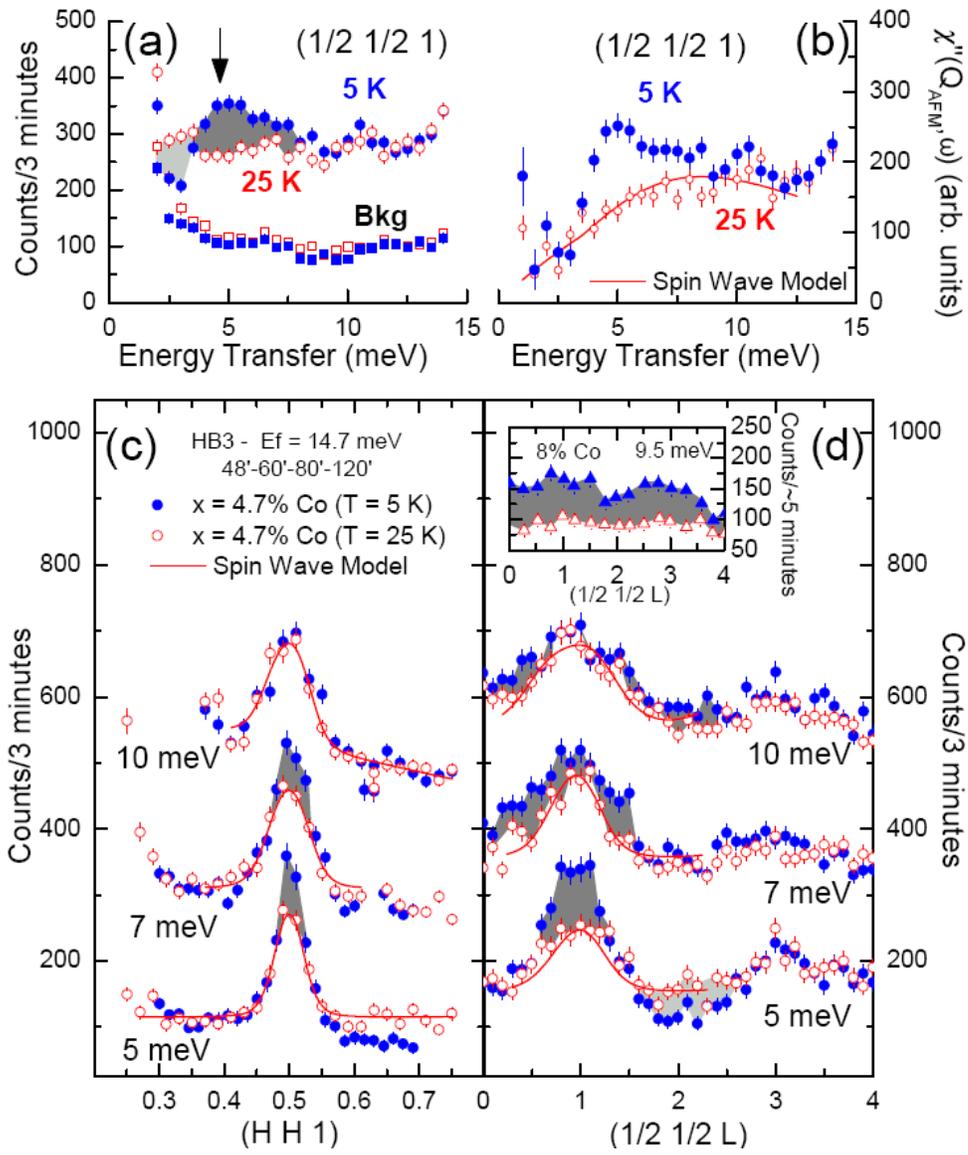

FIG. 1.

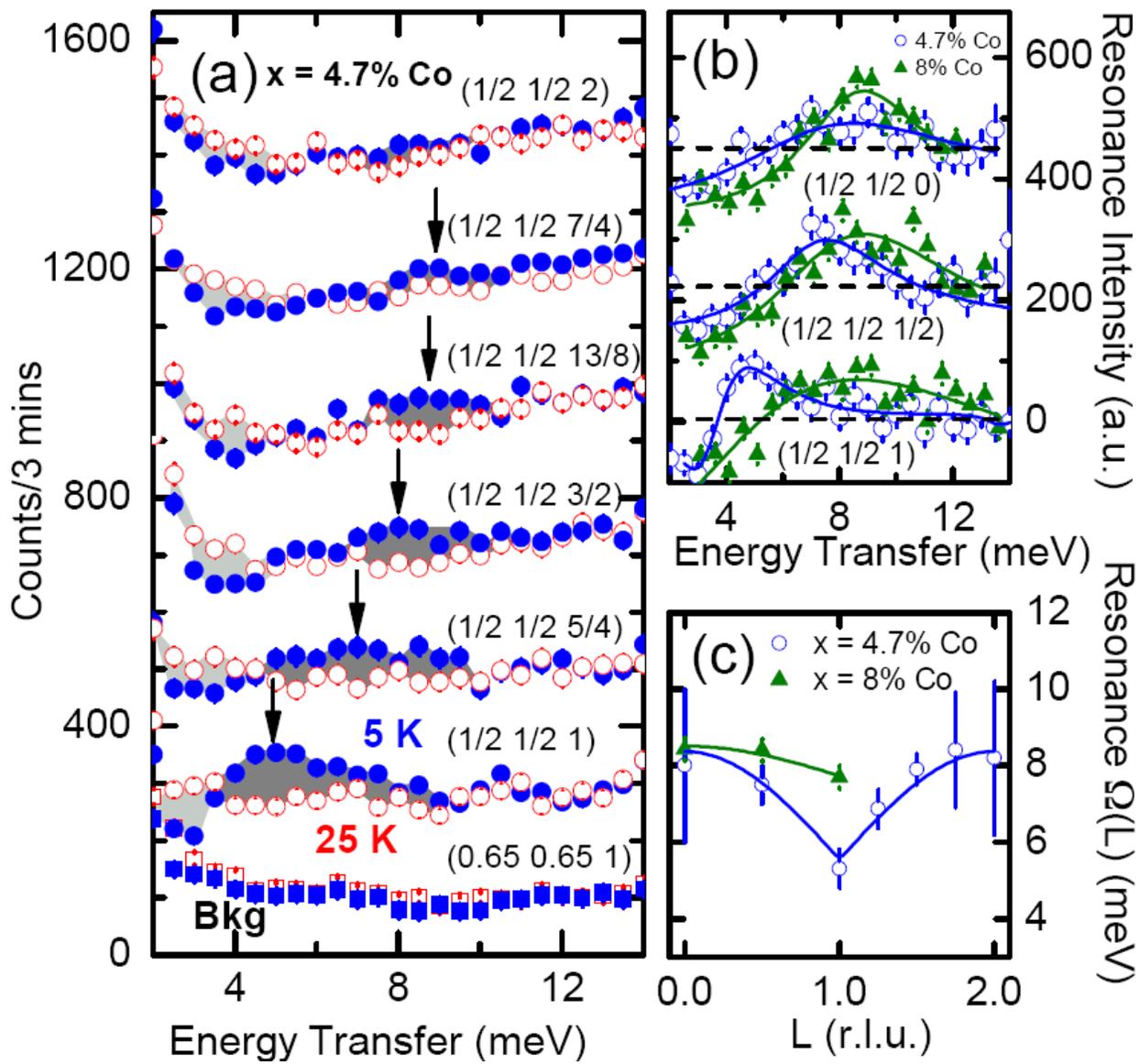

FIG. 2.

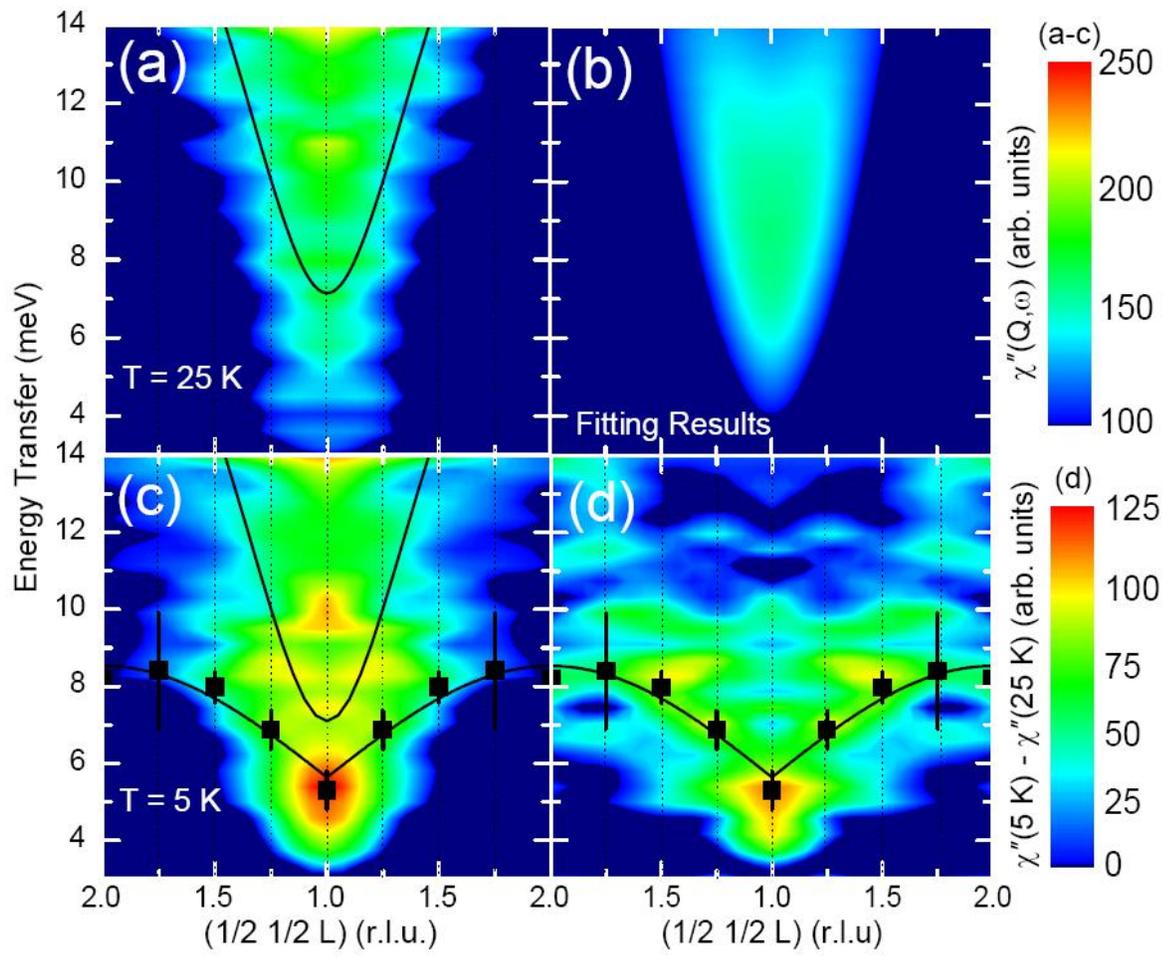

FIG. 3.

**REFERENCES**

[1] N. Ni *et al.*, Phys. Rev. B **78**, 214515 (2008).
[2] P. C. Canfield *et al.*, Phys. Rev. B **80**, 060501 (2009).
[3] N. Ni *et al.*, Phys. Rev. B **80**, 024511 (2009).
[4] J.-H. Chu *et al.*, Phys. Rev. B **79**, 014506 (2009).
[5] D. K. Pratt *et al.*, Phys. Rev. Lett. **103**, 087001 (2009).
[6] A. D. Christianson *et al.*, Phys. Rev. Lett. **103**, 087002 (2009).
[7] M. D. Lumsden *et al.*, Phys. Rev. Lett. **102**, 107005 (2009).
[8] S. Chi *et al.*, Phys. Rev. Lett. **102**, 107006 (2009).
[9] S. Li *et al.*, Phys. Rev. B **79**, 174527 (2009).
[10] D. S. Inosov *et al.*, Nat. Phys. **advance online publication** (2009).
[11] R. J. Birgeneau *et al.*, J. Phys. Soc. Japan **75**, 111003.
[12] M. Eschrig, Adv. Phys. **55**, 47 (2006).
[13] M. M. Korshunov and I. Eremin, Phys. Rev. B **78**, 140509 (2008).
[14] T. A. Maier and D. J. Scalapino, Phys. Rev. B **78**, 020514 (2008).
[15] Y. J. Uemura, Nat. Mater. **8**, 253 (2009).
[16] N. Bernhoeft *et al.*, Phys. Rev. Lett. **81**, 4244 (1998).
[17] A. Hiess *et al.*, J. Phys.: Condens. Matter **18**, R437 (2006).
[18] C. Stock *et al.*, Phys. Rev. Lett. **100**, 087001 (2008).
[19] A. Chubukov, D. Pines, and J. Schmalian, in *The Physics of Conventional and Unconventional Superconductors*, edited by K. H. Bennemann, and J. B. Ketterson (Springer-Verlag, Berlin, 2002), p. 495.
[20] J. P. Ismer *et al.*, Phys. Rev. Lett. **99**, 047005 (2007).
[21] J. Chang *et al.*, Phys. Rev. B **75**, 024503 (2007).
[22] A. V. Chubukov and L. P. Gor'kov, Phys. Rev. Lett. **101**, 147004 (2008).
[23] R. M. Fernandes *et al.*, arXiv:0911.5183 (2009).